\documentclass[twoside]{dis09}
\usepackage[latin1]{inputenc}
\usepackage[dvips]{graphicx,epsfig,color}
\usepackage{wrapfig,rotating}
\usepackage{amssymb,amsmath,array}

\begin{document}
\title{STAR heavy-flavor results}

\author{Jaroslav Bielcik for STAR collaboration
%
\thanks{This work was supported by grant INGO LA09013 of the Ministry of Education, Youth and Sports of the Czech
Republic.}
%
\vspace{.3cm}\\
%
Czech Technical University in Prague,\\ Faculty of Nuclear Sciences and Physical Engineering, \\Department of Physics\\
Brehova 7, Praha, 115 19, Czech Republic
}
%


\maketitle

\begin{abstract}
A summary of the heavy flavor results from the STAR experiment is presented. Both 
open heavy flavor as well as quarkonia measurements are presented.
A strong suppression of heavy flavor non-photonic  electrons is observed in central
Au+Au collisions at $\sqrt{s_{NN}}=$~200GeV. Relative contribution of bottom contribution to non-photonic electron
spectra in p+p collisions is extracted from data . Nuclear modification factor of J/$\Psi$ mesons at high-$p_{T}$
is found to be consistent with one in central Cu+Cu collisions  at $\sqrt{s_{NN}}=$~200GeV. Strong signal of 
$\Upsilon$(1S+2S+3S) state is observed in d+Au collisions at $\sqrt{s_{NN}}=$~200GeV.   
  
\end{abstract}

\section{Introduction}

   The calculation of Quantum Chromodynamics (QCD) on lattice showed that under conditions 
of high energy density or  high temperature nuclear matter undergoes a phase transition from 
state of confined quarks and gluons to deconfined state the Quark-Gluon Plasma (QGP). 
Such conditions were present in first moments after the Big Bang 
in the early universe and can be created in laboratory by colliding of heavy ions with sufficient energy. 
The results from experiments at the Relativistic Heavy Ion Collider (RHIC) at Brookhaven National Laboratory 
 provided plenty of 
information about this state of nuclear matter  during recent years \cite{Adams:2005dq}.  
   STAR measured a suppression of production of inclusive and also identified hadrons with high transverse momentum ($p_{T}$)
in central Au+Au collisions at $\sqrt{s_{NN}}=$~200GeV \cite{Adams:2006}. This is generally understood as a result of energy loss of energetic light partons in 
nuclear matter. Heavy quarks are believed to be mostly created through gluon fusion in the initial
phase of the collision and therefore they are well suited to provide and additional information about the parton energy loss 
and properties of QGP.  Due to  mass dependent suppression of gluon radiation under small angles, the dead-cone effect, it is expected that energy loss of
heavy quarks when compared with light partons should be significantly reduced 
\cite{Dokshitzer:2001zm}. 

 In following the STAR open heavy flavor and quarkonia measurements are discussed.

  \section{Open heavy flavor}

\begin{figure}[ht!]
\begin{minipage}{14pc}
\includegraphics[width=14pc]{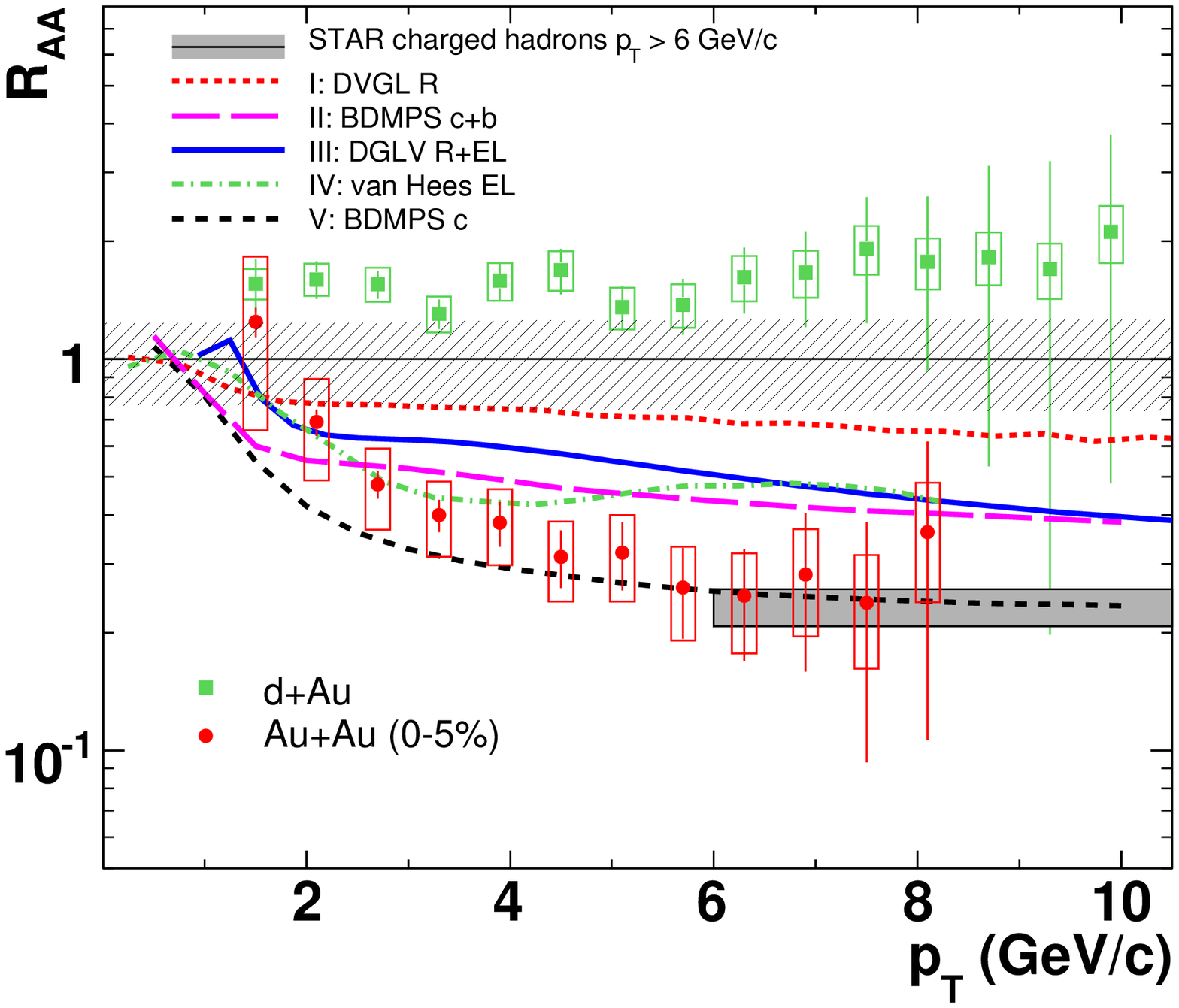}
\caption{\label{raa}Nuclear modification factor of non-photonic electrons in central Au+Au (circles) and d+Au  (squares) collisions at $\sqrt{s}$=200 GeV. Lines corresponds to theoretical calculations of energy loss.}
\end{minipage}\hspace{2pc}%
\begin{minipage}{14pc}
\includegraphics[width=14pc]{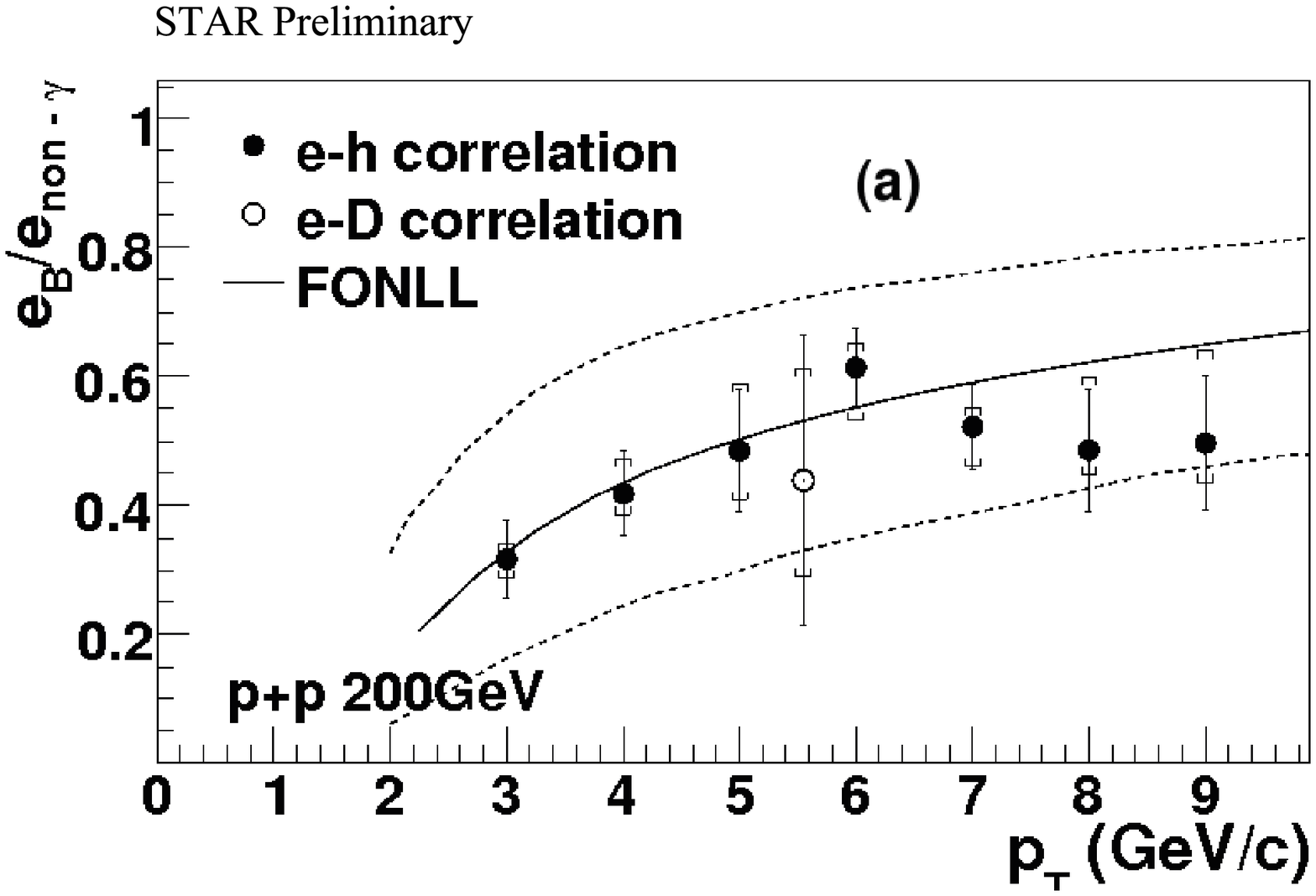}
\caption{\label{ratio} Relative contribution of bottom semileptonic electrons
to total non-photonic yield as a function of transverse momentum in p+p collisions at  $\sqrt{s}$=200 GeV. Lines correspond to FONLL pQCD prediction with lower and upper limit.}
\end{minipage} 
\end{figure}


   Open heavy flavor can be measured either directly by reconstruction of heavy flavor mesons (open charm D, open beauty B) 
or indirectly by measurement of non-photonic electrons or muons from its semileptonic decays. Direct reconstruction of 
heavy flavor mesons at RHIC is challenging due to  secondary vertex measurement limitation of current experiments. 
STAR \cite{Ackermann:2002ad} is able  to reconstruct directly   $D^0$ mesons in the hadronic channel $D^0\rightarrow K^-\pi^+$ using the information from 
Time Projection Chamber (TPC). This has been performed in several collision systems at 
$\sqrt{s_{NN}}$=200 GeV: d+Au \cite{STARcharmpaper}, Cu+Cu \cite{Baumgart:2008zz} and Au+Au \cite{:2008hja}. The transverse momentum spectrum  of  $D^0$ mesons  was extracted up 
to $p_{T}$ of ~3~GeV/$c$. The charm cross sections were extracted by
 a combined fit of $D^0$ and 
non-photonic electron and muon spectra. A binary scaling of charm cross section was observed 
in agreement with  expectation due to dominant charm production in hard scatterings \cite{Lin:1995pk}. 
     STAR reported measurements of high-$p_{T}$ non-photonic electron production in p+p,d+Au, 
Au+Au \cite{Abelev:2006db} and Cu+Cu collisions \cite{Knospe:2008zz}  at $\sqrt{s_{NN}}$=200 GeV. 
  For these measurements information from the TPC and the  Electro-Magnetic Calorimeter (EMC) were used.
The background contributions to the non-photonic electron spectrum from photonic sources, such as gamma  conversions or $\pi^{0}$ Dalitz decays, are statistically subtracted.  The EMC is also 
used in a trigger scheme during data taking to enhance the  high-$p_{T}$ part of the electron spectrum.  FONLL pQCD calculation predicts that from $p_{T}$ of 5~GeV/c, bottom decays contribute significantly to the non-photonic electron spectrum.

In Figure~\ref{raa}, the nuclear modification factor ($R_{AA}$)  of  non-photonic electrons is shown for d+Au and 5$\%$ most central Au+Au collisions. The $R_{AA}$ is 
the ratio of the yield measured in Au+Au 
to the yield from p+p collisions  - each scaled by the mean number of binary collisions 
in the collision system. In a case of d+Au collisions, observed $R_{AA}$ is close to one with a
possible small enhancement. However in the case of central Au+Au collisions a strong suppression
 by a  factor of $\sim$5 is observed. Note, that this is as strong as suppression of inclusive 
charged hadrons with $p_{T}>6$ GeV/c (shaded area). The  measured  $R_{AA}$ is predicted in
 several 
models  (see lines in Figure~\ref{raa}) \cite{Armesto:2005mz,Djordjevic:2005,Wicks:2005gt,vanHees:2005wb,adil2007}.  In general, the models over-predict the observed suppression 
of non-photonic electrons when the parameters of models (e.g. $dN_{g}/dy$) 
are constrained by the suppression of light hadrons. Only if the contribution of
bottom quarks is not taken into account, then the suppression is described well.

For further interpretation of the observed suppression it is crucial to address the 
contribution of bottom semileptonic electrons to measured non-photonic electrons spectra.
A study of the azimuthal correlation functions between non-photonic 
electrons and hadrons  was used to extract this information from data \cite{BiritzQM09}.  
Due to decay kinematics, there is a correlation between electrons and hadrons 
from semileptonic decays of charm and bottom mesons. Since the mass difference between charm and bottom mesons is large, there is also a large difference between the kinetic energy that mesons can give to the decay daughters. $e_{b}$/($e_{b}$ +$e_{d}$), the fraction of the total 
non-photonic electrons  due to B-meson decay, is plotted in Figure~\ref{ratio}. 
The measured value of the relative contribution of $e_{b}$/($e_{b}$ +$e_{d}$) 
is consistent with the FONLL predictions. This is an indication that the b-quark 
contribution  should be taken into account in describing and interpreting the 
suppression of non-photonic electrons already at moderate $p_{T}$.  

\section{Quarkonia measurements}

\begin{wrapfigure}{r}{0.5\columnwidth}
\centerline{\includegraphics[width=0.45\columnwidth]{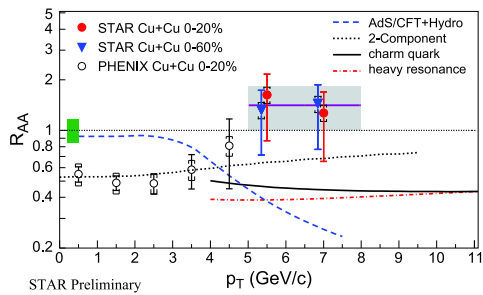}}
\caption{Nuclear modification factor of $J/\Psi$ in Cu+Cu collisions at$\sqrt{s_{NN}}$=200 GeV.
The solid line and band show the average and uncertainty of the two 0-20$\%$
data points. The curves are model calculations described in
the text. }\label{jpsi}
\end{wrapfigure}

      The main question in quarkonia studies we would like to address is how, possibly formed, 
QGP influences  the quarkonia yields. As result of the color Debye screening of the strong 
heavy quark  interaction in dense nuclear matter, the binding energy of a bound quarkonium
state decreases and it could lead to its disassociation \cite{Matsui:1986dk}. This would be 
manifested as suppression of quarkonia yields in heavy ion collisions when compared to 
p+p collisions. Since different quarkonia have different binding energy, they would 
disassociate at different temperature of QGP and therefore could be used as a thermometer.

     The suppression of $J/\Psi$ production was observed at the  CERN SPS experiments. Surprisingly
a suppression of similar strength was observed at RHIC for $J/\Psi$ with $p_{T} <$5~GeV/c at mid-rapidity, although the energy density and temperature reached at RHIC are much higher than at the SPS.  The suppression at RHIC  at forward rapidity is stronger than at mid-rapidity.    
This indicated that additional mechanisms influence the observed $J/\Psi$ yield, such as
recombination, feeddown from higher charmonia states, feeddown from B mesons. The hot wind dissociation model \cite{Liu} predicts that effective dissociation  temperature decreases
with increasing $J/\Psi$  velocity. Therefore the suppression is 
expected to be stronger at high $p_{T}$. 

     Figure~\ref{jpsi} shows the nuclear modification factor of $J/\Psi$ production at 
high-$p_{T}$ measured by STAR in $20\%$ and $60\%$ most central Cu+Cu collisions  at 
 $\sqrt{s_{NN}}$=200 GeV together with low $p_{T}$ measurement of the PHENIX collaboration. 
The averaged $R_{AA}$ value of two STAR points for centrality $0-20\%$ is $R_{AA}$ = 1.4 $\pm$ 0.4 (stat.) $\pm$ 0.2 (syst.), consistent with unity. $J/\Psi$ mesons are reconstructed
via $e^{+}e^{-}$ channel. Lines on the Figure show prediction
from theoretical models. The dashed curve shows the predictions of an AdS/CFT-based
calculation embedded in a hydrodynamic model with hot wind dissociation \cite{Liu} 
incorporated.  The predicted $p_{T}$ dependence is contrary to the measured data. The dotted line 
represents calculations of the  two component model \cite{Zhao:2008vu}.
This model includes $J/\Psi$  dissociation, statistical c$\bar{c}$  coalescence, formation time effects
and B-meson feeddown. The model describes the overall trend of the data fairly well.
The calculations represented by the solid and dash-dotted lines are based on models
WHDG and GLV  of charm energy loss. They qualitatively  describe the heavy flavor 
suppression  in Au+Au however predict strong $J/\Psi$ suppression for $p_{T} >$ 5GeV/c, 
contrary to the data. These results may suggest that the high-$p_{T}$ $J/\Psi$  production is 
dominated by the color singlet channel.

\begin{figure}[t]
\centerline{\epsfxsize=10cm\epsfbox{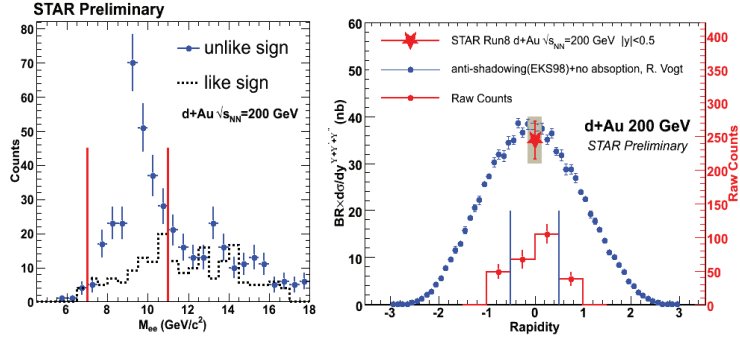}}   
\caption{(Left) Measurement of $\Upsilon \rightarrow e^{+}e^{-}$  signal and background 
in $\sqrt{s_{NN}}$ = 200GeV collisions. The solid symbols with statistical error bars 
are obtained by combining the unlike-sign ($e^{+}e^{-}$) pairs. The dashed histogram shows the like-sign background.(Right) The red star shows the measured $BR\times(\frac{d\sigma}{dy})^{\Upsilon(1S+2S+3S)}_{y=0}$ at midrapidity. 
The cross section is compared with the NLO CEM model prediction (blue solid circles),
see text for details. The raw yields vs. y are shown by the red histogram at the bottom
with the statistical errors.  \label{upsilon}}
\end{figure}

Another quarkonium system studied in $e^{+}e^{-}$ channel  in STAR is $\Upsilon$ meson.
Lattice QCD calculations predict  that $\Upsilon^{''}$ melts at RHIC energies, $\Upsilon^{'}$probably melts, however $\Upsilon$ state survives \cite{Digal:2001iu,Wong:2004zr}.
STAR has employed two level $\Upsilon$ trigger during data taking and reported measurements 
in p+p \cite{Djawotho:2007mj} and Au+Au collisions \cite{Das}. 

During the RHIC run 2008 STAR measured d+Au collisions at $\sqrt{s_{NN}}$=200 GeV. Due to
removal of  inner silicon detectors from the detector setup, lower background from 
$\gamma$ conversion electrons was present. Two levels of the STAR $\Upsilon$ trigger 
consist from hardware and software level. Only events with a BEMC tower with
deposited energy above the threshold passed Level 0. After clustering of towers in the  Level 2 
conditions on opening angle, deposited energy and invariant mass of possible $\Upsilon$
candidates were required. Details of the analysis steps are  presented in \cite{upsilonqm}.       
Figure~\ref{upsilon} (left) shows unlike sign and like sign combination of 
electron-positron pairs. In the region of expected $\Upsilon$ signal (horizontal lines)
very small background is populated. Strong signal  $\Upsilon + \Upsilon^{'}+\Upsilon^{''}$  
states of 8$\sigma$ significance and the integral count of 172 $\pm$ 20 ({\it stat.}) was 
obtained. The extracted value of the cross section to $e^{+}e^{-}$ channel is found as 
$BR\times(\frac{d\sigma}{dy})^{\Upsilon(1S+2S+3S)}_{y=0}$=35 $\pm$ 4(stat.) $\pm$ 5(sys.)nb.
at midrapidity in $\sqrt{s_{NN}}$ = 200GeV d+Au collisions. Figure~\ref{upsilon} (right) 
shows a comparison of the  measurement with the Color Evaporation Model NLO pQCD calculation \cite{Frawley:2008kk} including the anti-shadowing effect and no absorption. 
Using the STAR p+p measurement of cross section nuclear modification factor of
$R_{dAu}$=0.98 $\pm$ 0.32(stat.) $\pm$ 0.28(sys.) was found.

\section{Conclusions}

In this paper recent results of heavy flavor physics from the STAR collaboration
at RHIC at $\sqrt{s_{NN}}$ = 200GeV energy were presented. The strong suppression 
of non-photonic electrons is observed in
central Au+Au collisions. The inability of theoretical models to describe this 
satisfactorily raises the  question about the mechanisms of energy loss of heavy quarks 
in nuclear matter. The relative contribution of bottom decays is an important factor 
in interpreting the  measured data and was found to be consistent with FONLL pQCD 
predictions. Measurement of $J/\Psi$ production in central Cu+Cu collisions revealed that
at high-$p_{T}$ the spectra are consistent with no suppression. This is in contrary with 
theoretical expectations.  The measurement of  $\Upsilon$ production at mid-rapidity
in d+Au collisions was reported. The production cross section is consistent with the NLO 
calculations.


\begin{footnotesize}



%

\end{footnotesize}


\end{document}